\documentclass[aps,amssymb,showpacs,twocolumn]{revtex4}
\usepackage{amssymb}
\usepackage{graphicx}
\usepackage{color}
\usepackage{textcomp}
\usepackage{ulem}
\begin{document}

\title{ Effect of Quark Dimension Reduction on Goldstone Mode in Magnetic Field  }
\author{Shijun Mao$^1$ and Yuxuan Wang$^{1,2}$}
\affiliation{$^1$School of Science, Xi'an Jiaotong University, Xi'an 710049, China\\
$^2$School of Physics, Peking University, Beijing 100871, China}

\begin{abstract}
The meson static properties are investigated in Pauli-Villars regularized Nambu--Jona-Lasinio model in strong magnetic field. The quark dimension reduction leads to not only the magnetic catalysis effect on chiral symmetry restoration but also a sudden jump of the mass of the Goldstone mode at the Mott transition temperature.
\end{abstract}

\date{\today}
\maketitle

It is well known that, the quark dimension reduction in magnetic field leads to an increasing critical temperature of chiral restoration, namely the magnetic catalysis effect~\cite{mc1,mc2,mc3,rev1,rev2,rev3,rev4}. From the Goldstone theorem, the spontaneous breaking of a global symmetry implies the existence of Goldstone bosons. In two-flavor case, the neutral pion is identified as the Goldstone mode in the presence of a magnetic field. The pion properties such as its mass and decay constant play an important role in chiral dynamics~\cite{g1,g2}. For instance, the neutral mesons are potential for explaining the inverse magnetic catalysis~\cite{fukushima,mao} and delayed magnetic catalysis~\cite{mao2}.

The Nambu--Jona-Lasinio (NJL) model at quark level describes well the chiral symmetry breaking in vacuum and its restoration at finite temperature and baryon density~\cite{njl1,njl2,njl3,njl4,njl5}. In the model, mesons are treated as quantum fluctuations, and neutral mesons can be affected by the external magnetic field through their constituent quarks. The neutral mesons in magnetized NJL model are investigated in vacuum and at finite temperature by taking different methods like the assumption of four-momentum independent meson polarizations~\cite{njl2}, magnetic field independent regularization scheme~\cite{mfir,mfir2}, and derivative expansion~\cite{ritus5,ritus6} and $\Phi$-derivable approach~\cite{phi}. In this paper, we focus on how the quark dimension reduction in magnetic field affects the neutral meson properties at finite temperature and density in the Pauli-Villars regularization scheme.

The SU(2) NJL model is defined through the Lagrangian density~\cite{njl1,njl2,njl3,njl4,njl5}
\begin{equation}
\label{njl}
{\cal L} = \bar{\psi}\left(i\gamma_\mu D^\mu+\mu \gamma_0-m_0\right)\psi+\frac{G}{2}\left[\left(\bar\psi\psi\right)^2+\left(\bar\psi i\gamma_5\tau\psi\right)^2\right],
\end{equation}
where the covariant derivative $D^\mu=\partial^\mu-iQ A^\mu$ couples quarks to the external magnetic field ${\bf B}=(0, 0, B)$ in $z$-direction, $Q=diag (Q_u,Q_d)=diag (2e/3,-e/3)$ and $\mu=diag (\mu_u,\mu_d) = diag (\mu_B/3, \mu_B/3)$ are electric charge and quark chemical potential matrices in flavor space with $\mu_B$ being baryon chemical potential, $G$ is the coupling constant in scalar and pseudo-scalar channels, and $m_0$ is the current quark mass characterizing the explicit chiral symmetry breaking.

Taking the Leung-Ritus-Wang method~\cite{ritus1,ritus2,ritus3,ritus4,ritus7}, the quark condensate $\langle\bar\psi\psi\rangle$ or the dynamical quark mass $m_q=m_0-G\langle\bar\psi\psi\rangle$ at mean field level is controlled by the gap equation~\cite{mc1,mc2,mc3,rev1,rev2,rev3,rev4}
\begin{equation}
\label{gap}
m_q(1-GJ_1)=m_0,
\end{equation}
with $J_1 = N_c\sum_{f,n}\alpha_n |Q_f B|/(2\pi) \int d p_z/(2\pi)J_0/(2 E_f)$ and $J_0=\tanh[E^+_f/(2T)]+\tanh[E^-_f/(2T)]$, where $N_c=3$ is the number of colors which is trivial in the NJL model, $\alpha_n=2-\delta_{n0}$ is the spin degeneracy, and $E^\pm_f=E_f\pm \mu_B/3$ are the quark energies with $E_f=\sqrt{p^2_z+2 n |Q_f B|+m_q^2}$. Note that, the quark three-momentum integration in vacuum is reduced to a summation over Landau energy levels plus a one-dimensional momentum integration in magnetic field. It is the quark dimension reduction that leads to the magnetic catalysis effect on chiral restoration.

In the NJL model, mesons are treated as quantum fluctuations above the mean field and constructed through random phase approximation (RPA)~\cite{njl2,njl3,njl4,njl5,zhuang}. Without magnetic field, the isospin triplet $\pi_0$ and $\pi_\pm$ and isospin singlet $\sigma$ are respectively the Goldstone modes and Higgs mode corresponding to the spontaneous breaking of chiral symmetry. Turning on the external magnetic field, only $\pi_0$ is the Goldstone mode. While all the quarks and mesons contribute to the thermodynamics of the quark-meson plasma, $\pi_0$, as the Goldstone mode, controls the thermodynamics in the chiral breaking phase, and quarks dominate the thermodynamics in the chiral restoration phase. Considering the possible sizeable contribution from $\sigma$ around the chiral phase transition, we focus on the properties of $\pi_0$ and $\sigma$ in the following. With the RPA method, the meson propagator $D_M$ can be expressed in terms of the meson polarization function or quark bubble $\Pi_M$,
\begin{equation}
D_M(q)=\frac{G}{1-G\Pi_M(q)}.
\end{equation}
Considering the Lorentz symmetry breaking at finite temperature, the quark bubble and the meson propagator depend separately on $q_0$ and ${\bf q}$, $\Pi_M(q_0, {\bf q})$ and $D_M(q_0,{\bf q})$. The meson pole mass $m_M$ is defined through the pole of the propagator at zero momentum ${\bf q}={\bf 0}$,
\begin{equation}
\label{mmass}
1-G\Pi_M(m_M, {\bf 0})=0,
\end{equation}
and the quark-meson coupling constant is related to the residue at the pole~\cite{njl2,njl3,njl4,njl5,ritus5,ritus6,fukushima,mao,mao2},
\begin{equation}
\label{coupling}
g^{(\mu)}_M = \left(g^{\mu\mu} \frac{\partial\Pi_M (q)}{\partial q^2_\mu}\right)^{-1/2}\Bigg|_{q_0=m_M,\ {\bf q}={\bf 0}},
\end{equation}
with the space-time metric $g^{\mu\nu} = diag (1,-1,-1,-1)$. Note that, due to the special direction of the magnetic field ${\bf B}$, there is no more a uniform quark-meson coupling. In the case with ${\bf B}$ in the $z$-direction, we have $g^{(1)}_M=g^{(2)}_M \neq g^{(3)}_M=g^{(0)}_M$.

At zero momentum ${\bf q}={\bf 0}$, the polarization function for neutral mesons $\pi_0$ and $\sigma$ can be simplified as
\begin{equation}
\label{pi}
\Pi_M(\omega,{\bf 0}) = J_1-(\omega^2-\epsilon_M^2) J_2(\omega^2)
\end{equation}
with
\begin{equation}
\label{j2}
J_2(\omega^2) = -N_c\sum_{f,n}\alpha_n \frac{|Q_f B|}{2\pi} \int \frac{d p_z}{2\pi}{J_0\over 2 E_f (4 E_f^2-w^2)}
\end{equation}
and $\epsilon_{\pi_0}=0$ and $\epsilon_\sigma=2m$, and the quark-meson coupling constant $g^{(0)}_{M }$ can also be expressed in terms of the function $J_2$,
\begin{equation}
g^{(0)}_{M }=\left[-J_2(m^2_M)-(w^2-\epsilon_M^2) \frac{\partial J_2}{\partial w^2}\Big|_{w^2=m^2_{M}} \right]^{-1/2}.
\end{equation}
The pion decay constant $f_{\pi_0}$ for the Goldstone mode is defined through the vacuum to one-pion axial-vector matrix element~\cite{njl2},
\begin{equation}
i k_\mu f_{\pi_0}^{(\nu)} =\text{Tr}\int \frac{d^4 p}{(2\pi)^4}\gamma_\mu\gamma_5 \frac{\tau_3}{2} S(p+{k\over 2})g_{\pi_0}^{(\nu)} \gamma_5 \tau_3 S(p-{k\over 2}),
\end{equation}
where the trace is done in spin, color and favor spaces, and a straightforward calculation gives
\begin{equation}
\label{decay}
f^{(0)}_{\pi_0} = -m_q\ g^{(0)}_{\pi_0 }\ J_2(m_{\pi_0}^2).
\end{equation}

The above discussed chiral condensate $\langle\bar\psi\psi\rangle$ and the Goldstone and Higgs modes $\pi_0$ and $\sigma$ are charge neutral. They are affected by the external magnetic field through the constituent quarks. As a consequence, the equations for the condensate, meson mass, quark-meson coupling constant and pion decay constant are formally the same as that in vacuum~\cite{zhuang}, except for the replacement of the quark momentum integration $2N_f\int d^3{\bf p}/(2\pi)^3$ by $\sum_{f,n}\alpha_n |Q_f B|/(2\pi)\int dp_z/(2\pi)$.

Because of the four-fermion interaction, the NJL model is not a renormalizable theory and needs regularization. The magnetic field does not cause extra ultraviolet divergence but introduces discrete Landau levels and anisotropy in momentum space. For an anisotropic system, the Pauli-Villars regularization scheme can guarantee the law of causality~\cite{mao}. The three parameters in the NJL model, namely the current quark mass $m_0$, the coupling constant $G$ and the cutoff $\Lambda$ in the Pauli-Villars regularization, can be fixed by fitting the quark condensate $\langle\bar\psi\psi\rangle$, pion mass $m_\pi$ and pion decay constant $f_\pi$ in vacuum. The fixed parameters are $G=3.44$ GeV$^{-2}$ and $\Lambda=1127$ MeV with $m_0=5$ MeV.

\begin{figure}[hbt]
\centering
\includegraphics[width=7cm]{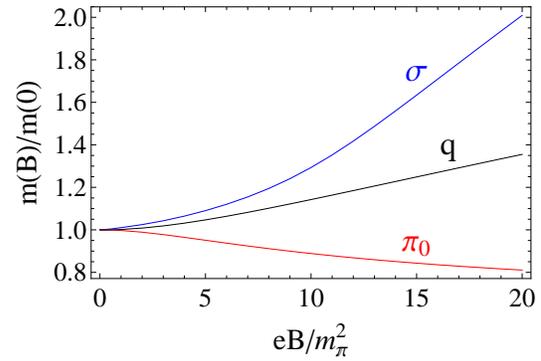}
\caption{ The mass ratio $m(B)/m(0)$ as a function of scaled magnetic field for quark and mesons in vacuum. }
\label{fig1}
\end{figure}

The magnetic field dependence of the quark and meson masses in vacuum is shown in Fig.\ref{fig1}. The quark mass $m_q$ and in turn the sigma mass ($m_\sigma \sim 2m_q$) increase with the strength of the magnetic field. On the other hand, as the pseudo-Goldstone mode, the pion mass $m_{\pi_0}$ decreases. These behaves are resulted from the quark dimension reduction mentioned above and consistent with the results of the NJL model with MFIR regularization scheme~\cite{mfir}, the chiral perturbation theory~\cite{hadron1} and lattice QCD simulation~\cite{l1,l2,l3}.

\begin{figure}[hbt]
\centering
\includegraphics[width=11cm]{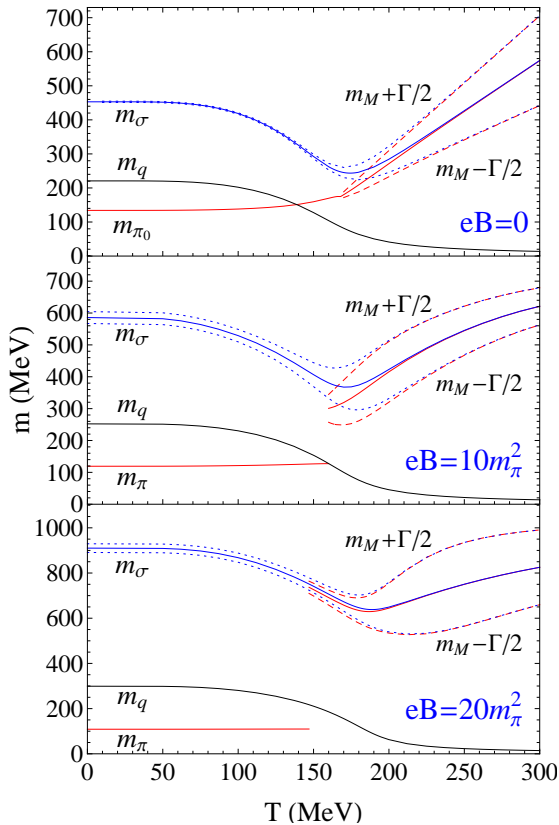}
\caption{ The quark and meson masses $m$ (solid lines) and widths $m_M\pm \Gamma/2$ (dashed lines) as functions of temperature at $\mu_B=0$ and $eB/m_\pi^2=0$, 10 and 20. }
\label{fig2}
\end{figure}

The temperature dependence of the meson masses is indicated in Fig.\ref{fig2} at $eB/m_\pi^2 = 0$, 10 and 20. We first consider the low and high temperature regions, corresponding to the chiral symmetry breaking and restoration phases. At low temperature, due to the gradual melting of the chiral condensate in medium, the quark mass, as the order parameter of chiral phase transition, drops down monotonically, while the pion mass slightly goes up with temperature but satisfies the bound state condition $m_{\pi_0}< 2m_q$. At high temperature with a small quark mass, the two neutral meson masses coincide and increase with temperature. When the pion mass is beyond the threshold $m_{\pi_0} > 2m_q$, the decay channel $\pi_0\to q + \bar q$ opens, and $\pi_0$ is no longer a bound state but a resonant state. In this case, the pole equation (\ref{mmass}) should be regarded in its complex form,
\begin{equation}
1-G\Pi_M \left(m_M-i \Gamma_M /2,{\bf 0} \right)=0,
\end{equation}
to determined the resonant mass $m_M$ and associated width $\Gamma_M$~\cite{zhuang}. Taking the approximation of small width in comparison with the mass, $m_M$ and $\Gamma_M$ are decoupled and we have
\begin{equation}
\Gamma_{M} \simeq {1\over m_{M}}\text {Im} \frac{1-G J_1}{G  J_2((m_{M}-i\epsilon)^2)}.
\end{equation}

Different from the Goldstone mode, the scalar meson $\sigma$ is always in resonant state in the whole temperature region, satisfying the decay condition $m_\sigma > 2m_q$. With increasing temperature, $m_\sigma$ decreases in the chiral breaking phase, reaches the minimum around the critical temperature, and then increases in the chiral restoration phase. Since the $\sigma$ width is very small at low temperature, we checked the $\pi_0-\sigma$ mass relation $m^2_\sigma=4m^2_q + m^2_{\pi_0}$ in the chiral breaking phase. It is slightly broken but the deviation is enlarged by the magnetic field.

We now turn to the meson masses around the critical temperature $T_c$ of chiral phase transition. When chiral symmetry is explicitly broken with a nonzero current quark mass $m_0$, there is no strict definition for the chiral symmetry restoration, and the critical temperature $T_c$ is normally defined as the temperature where the quark mass has the maximum change, $\partial^2 m_q(T)/\partial T^2|_{T_c}=0$. From Fig.\ref{fig2}, we have numerically $T_c=157$, 164 and 180 MeV at the magnetic field $eB/m_\pi^2=0$, 10 and 20, respectively. The increase of $T_c$ with $B$ is the typical result of mean field calculation in effective models, namely the magnetic catalysis\cite{mc1,mc2,mc3,rev1,rev2,rev3,rev4} induced by the quark dimension reduction in magnetic field. The other consequence of this dimension reduction is the mass jump for the Goldstone mode at the Mott transition point $T_m$ where the pion mass suddenly jumps up from $m_{\pi_0} < 2m_q$ to $m_{\pi_0} > 2m_q$. Looking back to the pole equation (\ref{mmass}), the jump arises from the singularity of the polarization function $\Pi_{\pi_0}(\omega,{\bf 0})$ at $\omega= 2 m_q$. From the explicit expression of $\Pi_M$, see (\ref{pi}) and (\ref{j2}), the factor $1/(4E_f^2-\omega^2)$ in the integrated function in $J_2$ becomes $(1/4)/(p_z^2+2n|Q_f B|)$ at $\omega= 2 m_q$. When we do the integration over $p_z$, the $p_z^2$ in the denominator leads to the infrared divergence at the lowest Landau level $n=0$. Therefore, $m_{\pi_0}= 2 m_q$ is not a solution of the pole equation, and there must be a mass jump for the Goldstone mode between its bound state with $m_{\pi_0} < 2m_q$ and $\Gamma_{\pi_0}=0$ at $T<T_m$ and resonant state with $m_{\pi_0} > 2m_q$ and $\Gamma_{\pi_0}\neq 0$ at $T>T_m$. From our numerical calculation, the temperature $T_m$ is 167, 160 and 147 MeV corresponding to the magnetic field $eB/m_\pi^2=0$, 10 and 20. Since $T_c$ increases with $B$ and $T_m$ decreases with $B$, the two temperatures should meet at some magnetic field. From our calculation, the cross point is located at $eB/m_\pi^2=5$. Note that, while the location of the jump and the relation between the critical temperature $T_c$ and Mott transition temperature $T_m$ depend on the model parameters and the definition of $T_c$, the mass jump is a direct result of the quark dimension reduction. When the magnetic field disappears, there is no more quark dimension reduction, the integration
$\int d^3{\bf p}/(4E_f^2-\omega^2)\sim \int dp$ becomes finite at $\omega = 2m_q$, and there is no more such a mass jump, see the upper panel of Fig.\ref{fig2}. It is also necessary to point out that, in hadron models like chiral perturbation theory and linear sigma model where hadrons are taken as elementary particles, the mass of the Goldstone mode changes with temperature continuously~\cite{hadron1,hadron2}.

\begin{figure}[hbt]
\centering
\includegraphics[width=8cm]{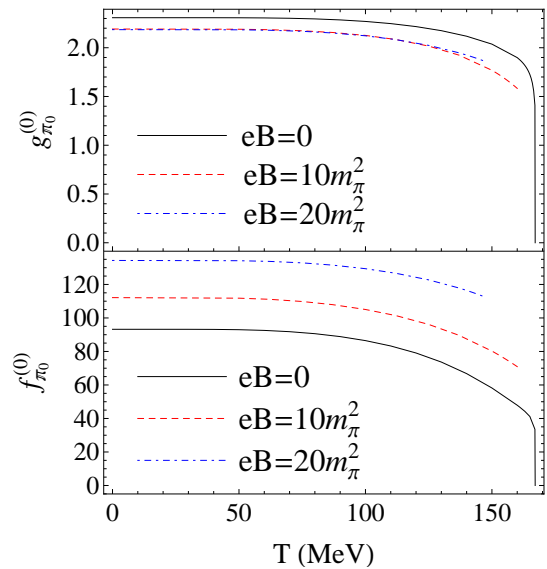}
\caption{The coupling constant $g_{\pi_0}^{(0)}$ and decay constant $f_{\pi_0}^{(0)}$ as functions of temperature at $\mu_B=0$ and $eB/m_\pi^2=0$, 10 and 20. }
\label{fig3}
\end{figure}

The coupling constant $g_{\pi_0}^{(0)}$ and decay constant $f^{(0)}_{\pi_0}$ for the Goldstone mode are depicted in Fig.\ref{fig3} as functions of temperature at different magnetic field $eB/m^2_\pi=0$, $10$ and $20$. In the beginning, both the coupling and decay constants are almost temperature independent, due to the slight change of the quark and $\pi_0$ masses shown in Fig.\ref{fig2}. Only when approaching to the Mott transition temperature $T_m$, they vary dramatically. At vanishing magnetic field they drop down to zero at $T_m$ very rapidly but continuously. When the field is turned on, however, the continuity at $T_m$ is replaced by a suddenly jump, arising from the mass jump shown in Fig.\ref{fig2}. At high temperature with $T > T_m$, mesons become resonant states, and the coupling and decay constants vanish.

At quark level, the Goldberger-Treiman relation and Gell-Man--Oakes--Renner relation are written as $(f^{(0)}_{\pi_0})^2 ( g_{\pi_0}^{(0)} )^2=m_q^2$ and $m^2_{\pi_0} (f^{(0)}_{\pi_0} )^2=m_0 m_q/G$~\cite{njl2}. We numerically checked these relations at finite temperature and magnetic field in the chiral breaking phase. While the two relations are well satisfied and almost independent of the field strength at low temperature, they are clearly broken when the system approaches to the Mott transition point.

\begin{figure}[hbt]
\centering
\includegraphics[width=11cm]{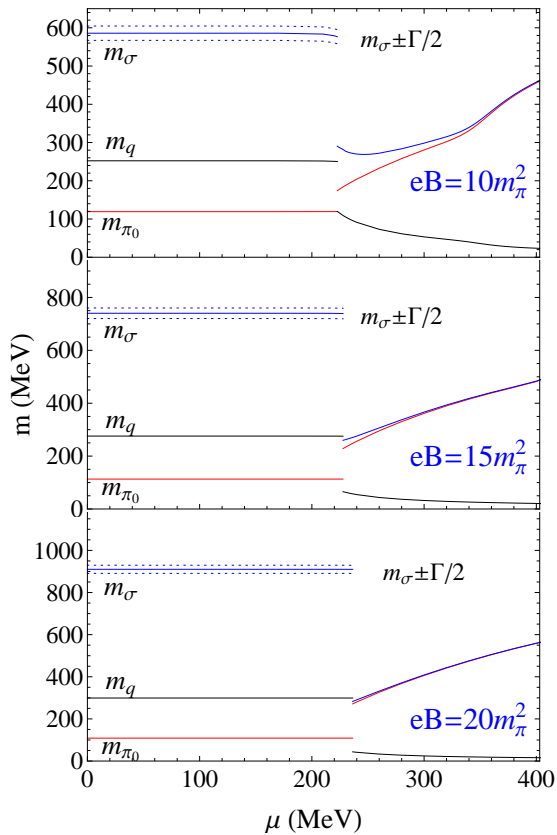}
\caption{The quark and meson masses (solid lines) and widths (dashed lines) as functions of baryon chemical potential at $T=0$ and $eB/m_\pi^2=10$, 15 and 20. }
\label{fig4}
\end{figure}

It is well known that, with increasing baryon chemical potential $\mu_B$ the chiral symmetry restoration changes from a crossover to a first order phase transition. What is the magnetic field effect in this case? In Fig.\ref{fig4} we show the quark and meson masses as functions of chemical potential at different magnetic field. It is easy to understand that, all the masses are constants below the Fermi surface. At the critical point $\mu_B^c$, the quark mass jumps down, indicating the chiral phase transition of the first order. Moreover, the mass jumps for the Goldstone and Higgs modes here are the consequence of the first order phase transition and not related to the Mott transition.

In fact, there is no more Mott transition at finite baryon chemical potential. Considering the Pauli blocking effect, the threshold for a meson to decay into a pair of quark and antiquark becomes~\cite{njl2,zhuang}
\begin{equation}
m_M > 2m_q\ \ \  \text{and}\ \ \  m_M > 2\mu_q=2/3\mu_B.
\end{equation}
The second condition means that the energy of quarks into which the meson decays have to be above the Fermi surface. In this case, only $\sigma$ satisfies both conditions in the chiral breaking phase at low chemical potential, $\pi_0$ at any chemical potential and $\sigma$ at $\mu_B > \mu_B^c$ are in bound states, see Fig.\ref{fig4}.

We now briefly discuss the meson properties in chiral limit with vanishing current quark mass $m_0=0$. Let's consider $T\neq 0$ and $\mu_B=0$. Comparing the gap equation $m_q(1-GJ_1)=0$ for quarks with the pole equation $1-G\Pi_M(m_M,{\bf 0})=0$ for mesons, the chiral phase transition is well defined and we have the analytic solutions
\begin{equation}
m_{\pi_0}=0,\ \ m_\sigma=2m_{q} \ \ \text{for}\ m_q\neq 0,
\end{equation}
in the chiral breaking phase and
\begin{equation}
m_\sigma=m_{\pi_0}\ \ \text{for}\ m_q=0
\end{equation}
in the chiral restoration phase. A direct consequence of these solutions is that the Mott transition temperature $T_m$ defined by $m_{\pi_0}(T_m)=2m_q(T_m)$ coincides with the critical temperature $T_c$ determined by $m_q(T_c)=0$. From the explicit expression (\ref{pi}) for the pion polarization function $\Pi_M$, while the integration over the quark momentum is still divergent at the critical temperature, the factor $(m_{\pi_0}^2-\epsilon_{\pi_0}^2)\to 0$ at this point cancels the divergence. As a result, the meson mass $m_{\pi_0}=2m_q$ is a solution of the pole equation at $T_c=T_m$, and the jump induced by the quark dimension reduction disappears in chiral limit.

In summary, the Goldstone mode ($\pi_0$) and Higgs mode ($\sigma$) of chiral symmetry breaking are investigated in a Pauli-Villars regularized NJL model at finite magnetic field. As quantum fluctuations above the mean field at quark level, the neutral mesons are affected by the external magnetic field through charged constituent quarks. The quark dimension reduction in magnetic field leads to not only the magnetic catalysis effect on chiral phase transition at mean field level but also a sudden mass jump for the Goldstone mode at the Mott transition point. As a consequence of such a jump, it may result in some interesting phenomena in relativistic heavy ion collisions where strong magnetic field can be created. For instance, when the formed fireball cools down, there might be a sudden enhancement of neutral pions at the Mott transition temperature.

\noindent {\bf Acknowledgement:}
The work is supported by the NSFC Grant 11405122 and China Postdoctoral Science Foundation Grant 2014M550483.

\end{document}